\documentclass{aa}

\usepackage{graphicx}
\usepackage[varg]{txfonts}
\usepackage[utf8]{inputenc}
\usepackage[T1]{fontenc}
\usepackage[normalem]{ulem}

\graphicspath{{./}{figures/}}

\title{Electron heat flux in the near-Sun environment}
\author{J. S. Halekas\inst{1}
\and P. L. Whittlesey\inst{2}
\and D. E. Larson\inst{2}
\and D. McGinnis\inst{1}
\and S. D. Bale\inst{2,3,4,5}
\and M. Berthomier\inst{6}
\and A. W. Case\inst{7}
\and B. D. G. Chandran\inst{8,9}
\and J. C. Kasper\inst{7,10}
\and K. G. Klein\inst{11}
\and K. E. Korreck\inst{7}
\and R. Livi\inst{2}
\and R. J. MacDowall\inst{12}
\and M. Maksimovic \inst{13}
\and D. M. Malaspina\inst{14,15}
\and L. Matteini \inst{16}
\and M. P. Pulupa\inst{2}
\and M. L. Stevens\inst{7}}

\institute{Department of Physics and Astronomy, University of Iowa, Iowa City, IA 52242, USA
\and Space Sciences Laboratory, University of California, Berkeley, CA 94720, USA
\and Physics Department, University of California, Berkeley, CA 94720, USA
\and The Blackett Laboratory, Imperial College London, London, SW7 2AZ, UK
\and School of Physics and Astronomy, Queen Mary University of London, London E1 4NS, UK
\and Laboratoire de Physique des Plasmas, CNRS, Sorbonne Universite, Ecole Polytechnique, Observatoire de Paris, Universite Paris-Saclay, Paris, 75005, France
\and Smithsonian Astrophysical Observatory, Cambridge, MA 02138, USA
\and Department of Physics \& Astronomy, University of New Hampshire, Durham, NH 03824, USA
\and Space Science Center, University of New Hampshire, Durham, NH 03824, USA
\and Climate and Space Sciences and Engineering, University of Michigan, Ann Arbor, MI 48109, USA
\and Department of Planetary Sciences, University of Arizona, Tucson, AZ 85721, USA
\and NASA Goddard Space Flight Center, Greenbelt, MD 20771, USA
\and LESIA, Observatoire de Paris, Universite PSL, CNRS, Sorbonne Universite, Universite de Paris, 5 place Jules Janssen, 92195 Meudon, France
\and Astrophysical and Planetary Sciences Department, University of Colorado, Boulder, CO 80309, USA
\and Laboratory for Atmospheric and Space Physics, University of Colorado, Boulder, CO 80303, USA
\and Physics Department, Imperial College London, London, SW7 2AZ, UK.
}

\date{Received 25 August 2020; Accepted 6 October 2020}

\begin{document}

\abstract
{}
{We survey the electron heat flux observed by the Parker Solar Probe (PSP) in the near-Sun environment at heliocentric distances of 0.125--0.25 AU.}
{We utilized measurements from the Solar Wind Electrons Alphas and Protons and FIELDS experiments to compute the solar wind electron heat flux and its components and to place these in context.}
{The PSP observations reveal a number of trends in the electron heat flux signatures near the Sun. The magnitude of the heat flux is anticorrelated with solar wind speed, likely as a result of the lower saturation heat flux in the higher-speed wind. When divided by the saturation heat flux, the resulting normalized net heat flux is anticorrelated with plasma beta on all PSP orbits, which is consistent with the operation of collisionless heat flux regulation mechanisms. The net heat flux also decreases in very high beta regions in the vicinity of the heliospheric current sheet, but in most cases of this type the omnidirectional suprathermal electron flux remains at a comparable level or even increases, seemingly inconsistent with disconnection from the Sun. The measured heat flux values appear inconsistent with regulation primarily by collisional mechanisms near the Sun. Instead, the observed heat flux dependence on plasma beta and the distribution of suprathermal electron parameters are both consistent with theoretical instability thresholds associated with oblique whistler and magnetosonic modes.}
{} 


\maketitle

\section{Introduction}
\label{sec:intro}

Solar wind electrons have complex velocity distribution functions (VDFs) that evolve with heliocentric distance under the influence of competing physical processes, including Lorentz and gravitational forces, Coulomb collisions, and plasma instabilities. The electron VDFs therefore carry information about these processes, with implications for the solar wind energy balance and the physics of the near-Sun environment. 

Large scale electric fields enforce the quasi-neutrality (equal electron and ion densities) and zero current (equal electron and ion charge fluxes) conditions on macroscopic scales. Nonzero electric fields naturally arise as a result of the different gravitational forces on electrons and ions \citep{pannekoek_ionization_1922}, leading to a significant potential drop from the corona. This electric potential may play a role in accelerating the solar wind ions \citep{lemaire_kinetic_1971, lemaire_kinetic_1973, scudder_causes_1992, pierrard_lorentzian_1996, maksimovic_kinetic_1997}, trapping a portion of the electron VDF \citep{boldyrev_electron_2020}, or driving a portion of the electron VDF into runaway \citep{scudder_steady_2019}. 

Meanwhile, electron gyromotion in diverging solar magnetic fields results in a mirror force that focuses the gyrating electrons. Absent of other influences, this would result in VDFs "beamed" along the magnetic field. However, Coulomb collisions \citep{scudder_theory_1979, scudder_theory_1979-1, salem_electron_2003, stverak_electron_2008, horaites_electron_2019, boldyrev_kinetic_2019} and plasma instabilities \citep{kennel_limit_1966, hollweg_new_1970, gary_heat_1975, gary_whistler_1994, gary_electron_1999, vocks_electron_2005, stverak_electron_2008, landi_competition_2012, roberg-clark_suppression_2018, vasko_whistler_2019, verscharen_self-induced_2019, innocenti_collisionless_2020} also act upon the electron VDFs, typically driving them toward isotropy. 

The electron VDFs that result from these competing influences have significant departures from an isotropic Maxwellian. Solar wind VDFs typically consist of a thermal "core" population with an approximately bi-Maxwellian form, a suprathermal "halo" also with an approximately bi-Maxwellian form but with a higher temperature, and a suprathermal magnetic field-aligned component known as the "strahl" that streams outward from the Sun \citep{feldman_solar_1975, rosenbauer_survey_1977, pilipp_characteristics_1987, maksimovic_radial_2005, stverak_radial_2009}. To maintain a balance in current, the core has a sunward drift with respect to the solar wind protons in order to balance the antisunward strahl as well as any drift of the halo \citep{feldman_solar_1975, scime_regulation_1994}. The VDF components evolve with heliocentric distance, with the fractional density of the strahl decreasing and that of the halo increasing  \citep{maksimovic_radial_2005, stverak_radial_2009, halekas_electrons_2020}, and the angular width of the strahl increasing \citep{hammond_variation_1996, graham_evolution_2017, bercic_scattering_2019}. These characteristics appear consistent with a scenario wherein the strahl represents a nearly free-streaming population escaping from the corona, with the halo generated from it by collisions or instabilities \citep{landi_competition_2012, horaites_electron_2019, boldyrev_kinetic_2019, bercic_coronal_2020}. However, the strahl and the halo could both instead ultimately arise from a sunward-directed runaway population \citep{scudder_steady_2019}. 

Whatever the cause(s) of the prevailing characteristics discussed above, they result in significant asymmetry and nonmaxwellianity of the electron VDFs. In other words, the VDFs have large values of both skew and kurtosis, and they therefore carry a large magnetic field-aligned electron heat flux $Q_{e_{||}} = \int{\frac{1}{2}mv^2v_{||}f(v)d^3v}$, where $||$ is along the magnetic field direction. This integral is taken in the solar wind frame, and so all three components of the electron VDF can contribute (positively or negatively) to the net electron heat flux.

The electron heat flux has a physical importance for several reasons. Electrons efficiently conduct heat, and the electron heat flux therefore contributes to the energy balance in the corona and the solar wind \citep{hollweg_electron_1974, hollweg_collisionless_1976, cranmer_empirical_2009, bale_electron_2013, landi_electron_2014, stverak_electron_2015}, as well as in other astrophysical contexts such as clouds within supernova remnants, galaxy clusters, and accretion flows \citep{cowie_evaporation_1977, bertschinger_role_1986, johnson_effects_2007, ressler_electron_2015, roberg-clark_suppression_2016}. The electron heat flux has different limiting values in different contexts, with collisional effects limiting the heat conduction in some regimes \citep{spitzer_transport_1953, salem_electron_2003, bale_electron_2013, horaites_self-similar_2015}, but collisionless effects associated with plasma instabilities playing a dominant role in other regimes \citep{feldman_interplanetary_1974, gary_heat_1975, gary_whistler_1994, scime_regulation_1994, gary_electron_1999, gary_whistler_2000, saeed_electron_2017, shaaban_clarifying_2018, roberg-clark_suppression_2018, roberg-clark_wave_2018, lopez_alternative_2020}. The relative importance of collisional and collisionless heat flux regulation mechanisms in the solar wind remains in question, with some studies suggesting that the electron heat flux approaches the collisional Spitzer-H{\"a}rm limit for a substantial fraction of the time at 1 AU \citep{bale_electron_2013}, but other work raising doubt as to whether the radial evolution of the distribution follows the Spitzer-H{\"a}rm result \citep{landi_electron_2014}. 

The radial evolution of the electron heat flux provides some insight into the processes that act upon the electron VDF. A completely unregulated heat flux would fall off at the same rate as the magnetic field magnitude, that is to say as $\sim r^{-2}$ close to the Sun and as $\sim r^{-1}$ farther from the Sun \citep{scime_regulation_1994}. In contrast, a fully collisionally regulated heat flux would fall off much faster, as $\sim r^{-5} - r^{-4}$ \citep{scime_regulation_1994}. None of the observations to date appear consistent with either of these limits, with reported electron heat flux radial profiles of $\sim r^{-3} - r^{-2.4}$ \citep{pilipp_large-scale_1990, mccomas_solar_1992, scime_regulation_1994, stverak_electron_2015, halekas_electrons_2020}. 

Given the observed heat flux radial profiles, considerable attention has focused on collisionless heat flux regulation mechanisms. Ideally, the candidate mechanism(s) would both regulate the heat flux and account for the radial trends in the strahl fractional density and width and the halo fractional density. Candidates identified to date include a variety of electrostatic instabilities \citep{gary_electrostatic_1978, roberg-clark_wave_2018, lopez_alternative_2020}, quasi-parallel whistler instabilities (often termed the "whistler heat flux instability" or WHFI) \citep{gary_heat_1975, gary_whistler_1994, gary_electron_1999, gary_whistler_2000, saeed_electron_2017, roberg-clark_wave_2018, shaaban_clarifying_2018, lopez_particle--cell_2019, lopez_alternative_2020}, oblique magnetosonic and whistler instabilities \citep{vasko_whistler_2019, horaites_stability_2018, boldyrev_kinetic_2019, verscharen_self-induced_2019, lopez_alternative_2020}, and the firehose instability \citep{shaaban_clarifying_2018, innocenti_collisionless_2020}. While whistler mode waves consistent with the WHFI exist in the solar wind \citep{lacombe_whistler_2014, stansby_experimental_2016, tong_statistical_2019, tong_whistler_2019, jagarlamudi_whistler_2020}, doubt remains as to whether this instability can adequately regulate the heat flux \citep{kuzichev_nonlinear_2019, lopez_particle--cell_2019, verscharen_self-induced_2019}. Oblique instabilities therefore provide an appealing candidate to scatter the strahl and regulate the heat flux. Observations confirm that oblique whistler mode waves also occur in the solar wind, during time periods with conditions favorable for instability growth \citep{breneman_observations_2010, cattell_narrowband_2020}. 
A correlation between sunward electron core drift and high frequency waves observed by PSP during its early orbits may point to another example of instabilities that influence the near-Sun evolution of the electron heat flux \citep{malaspina_plasma_2020}.  

Since the net electron heat flux is reliably outward along magnetic field lines from the Sun, its characteristics are often utilized to diagnose magnetic field topology. For example, the observation of sunward electron heat flux \citep{owens_sunward_2017} generally indicates that magnetic field lines follow an S-shaped curve or "switchback", as ubiquitously observed within 0.25 AU of the Sun \citep{kasper_alfvenic_2019, bale_highly_2019}. Similarly, the presence of bidirectional electron heat flux indicates magnetic field lines connected to the Sun at both ends \citep{gosling_bidirectional_1987}. 

Conversely, dropouts in the electron heat flux may indicate magnetic field lines completely disconnected from the Sun, presumably by magnetic reconnection \citep{mccomas_electron_1989}. However, not all electron heat flux dropouts indicate disconnection \citep{lin_interplanetary_1992}, with some heat flux dropouts instead resulting from bidirectional or isotropic suprathermal electron fluxes \citep{pagel_assessing_2005, pagel_understanding_2005}. Periods with nearly isotropic suprathermal electron distributions exhibit reduced net heat flux but no drop in omnidirectional suprathermal flux. Such distributions appear to occur preferentially during periods with high plasma beta, for example the "heliospheric plasma sheet" (HPS) regions surrounding the heliospheric current sheet (HCS) and other magnetic field reversals \citep{winterhalter_heliospheric_1994, crooker_heliospheric_2004}, and may result from enhanced scattering processes \citep{crooker_suprathermal_2003}. 

In this manuscript we analyze the electron heat flux measured within 0.25 AU of the Sun by the Parker Solar Probe (PSP) \citep{fox_solar_2016}. We utilized observations from the first two PSP orbits, as well as the fourth and fifth  orbits which for the first time reached a heliocentric distance of 0.125 AU.  

\section{Parker Solar Probe electron measurements}
\label{sec:meas}

To determine the electron properties, we utilized charged particle measurements from the Solar Wind Electrons Alphas and Protons (SWEAP) experiment \citep{kasper_solar_2016}, including the Solar Probe ANalyzers-Electrons (SPAN-E) \citep{whittlesey_solar_2020}, the Solar Probe Cup (SPC) \citep{case_solar_2020}, and the Solar Probe ANalyzer-Ions (SPAN-I), as well as magnetic field measurements from the FIELDS experiment \citep{bale_fields_2016}. To determine the electron core density, parallel and perpendicular temperatures, and drift speed, we utilized the fitting methodology described in \citet{halekas_electrons_2020}. We simultaneously fit the core and a low energy population of secondary electrons from the spacecraft, in the solar wind proton frame, as determined by SPC (for the first two PSP orbits) or SPAN-I (for the fourth and fifth orbits). Other authors \citep{bercic_coronal_2020} have instead elected to simply remove the portion of the measurement below ~20-30 eV, with very comparable results. The SPAN-E relative sensitivity has been calibrated by enforcing gyrotropy in the solar wind frame at higher energies where no secondary electron contamination is present \citep{halekas_electrons_2020}, and the absolute sensitivity has been determined by comparing to results from quasithermal noise spectroscopy \citep{moncuquet_first_2020}. 

To characterize the suprathermal (noncore) electrons, we utilized numerically integrated moments of the suprathermal residual of the measured distribution with respect to the best-fit core model. Rather than explicitly separating the strahl and halo, we considered the suprathermal population as a whole; however, in some cases we separated the suprathermal electrons in the magnetic field-aligned and antimagnetic field-aligned halves of the VDF. In contrast to \citet{halekas_electrons_2020}, which utilized moment computations in 3-d, we utilized a 2-d integration over a gyro-averaged distribution constructed by binning the measurements from both sensors in twelve $15\degr$ pitch angle sectors. This method has the advantage of effectively "filling in" gaps in phase space, by exploiting the gyrotropy of the distribution in the solar wind frame. If the magnetic field rotates during the SPAN-E measurement, it will affect the binning into a 2-d distribution. At most times, this does not appear to adversely affect the results, as demonstrated by the largely well formed pitch angle distributions observed. Gaps in phase space coverage can still occur when the magnetic field direction lies in a hole in the field of view, resulting in an underestimation of the heat flux carried by the strahl for some field orientations (primarily the most radial fields), but this procedure minimizes such effects. We have not explicitly excluded such distributions from the following analyses, but we have checked at all stages to verify that their inclusion does not affect any of the conclusions. We limited the integration to electron energies above the core temperature, in order to avoid the omnipresent secondary electron contamination at low energy. 

\begin{figure}
    \resizebox{\hsize}{!}{\includegraphics{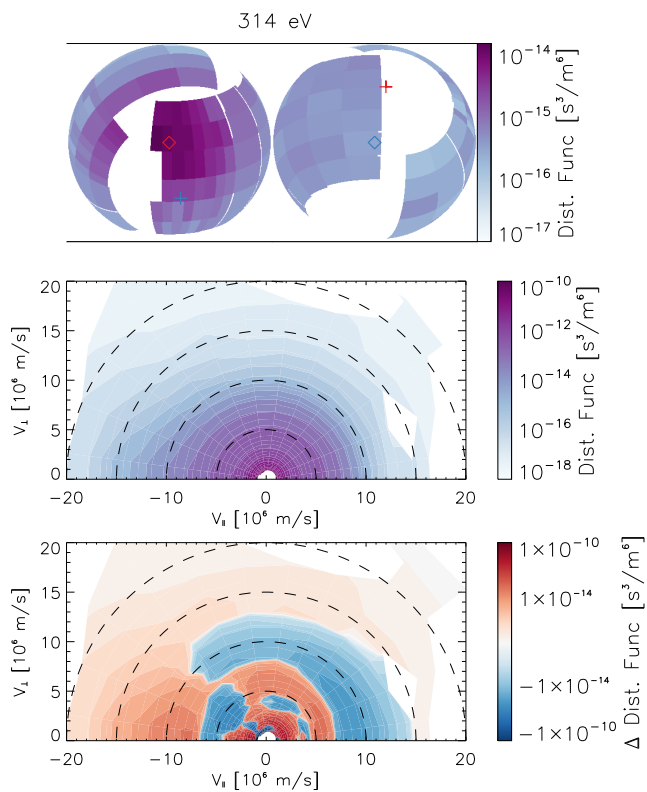}}
    \caption{Electron velocity distribution function (VDF) measured at 12:40:20 UT on 31 Jan 2020 by the Solar Wind Electrons Alphas and Protons (SWEAP) experiment on the Parker Solar Probe (PSP). The top panels show orthographic projections of the VDF at 314 eV (with blocked and/or contaminated pixels removed), in the solar wind proton frame. Symbols indicate electron velocities parallel (blue) and antiparallel (red) to the sunward magnetic field (diamonds) and the antisunward proton velocity (pluses). The center panel shows a 2-d projection of the same electron VDF, in magnetic-field aligned coordinates. The bottom panel shows the difference between the measured VDF and a drifting bi-Maxwellian fit to the electron core.}
    \label{fig:vdf}
\end{figure}

Figure \ref{fig:vdf} shows an illustrative example of our  methods. For this VDF, observed within the inward magnetic polarity sector on the fourth PSP orbit, we found a best-fit core distribution with a density of 466 $\mbox{cm}^{-3}$, parallel and perpendicular temperature components of 52 and 43 eV, and a magnetic field-aligned sunward core drift of +127 km/s. By performing a 2-d integration of the residual of the measured VDF with respect to the best-fit core model (bottom panel of Fig. \ref{fig:vdf}), we found a suprathermal electron heat flux of $-2.80 \times 10^{-3} \ \mbox{W/m}^2$. This heat flux is carried partly by the suprathermal surplus in the antimagnetic field direction ($-1.59 \times 10^{-3} \ \mbox{W/m}^2$), and partly by a suprathermal deficit in the magnetic field direction ($-1.22 \times 10^{-3} \  \mbox{W/m}^2$), both of which contribute to the skew of the VDF and thus to the net heat flux. We commonly observe a deficit (with respect to a Maxwellian) in the sunward-going portion of the VDF close to the Sun \citep{halekas_electrons_2020, bercic_coronal_2020}; this example has a particularly large deficit. Meanwhile, the core drift carries a heat flux of $1.14 \times 10^{-3} \  \mbox{W/m}^2$ in the magnetic field-aligned direction, as determined from the core fit parameters. To find the net electron heat flux, we summed all these terms to obtain $-1.66 \times 10^{-3} \ \mbox{W/m}^2$. The net heat flux thus involves a combination of fit-based results (the core) and numerical integrations (the suprathermal population). We find that the currents carried by these populations typically balance to satisfy the zero current condition in the plasma frame, as expected from physical principles, providing confidence in the overall accuracy of the scheme. 

We also grouped these three heat flux terms into components corresponding to the heat flux carried by the parallel and antiparallel (magnetic field-aligned and antimagnetic field-aligned) portions of the distribution, with the entire core heat flux included with the component corresponding to its sign (+ for parallel, $-$ for antiparallel). This provides more information than a single net heat flux number, especially in cases where there is significant heat flux in both directions along the field line, which cancels out in the net heat flux. In this case, we found parallel and antiparallel heat flux components of $-8 \times 10^{-5} \ \mbox{W/m}^2$ and $-1.59 \times 10^{-3} \ \mbox{W/m}^2$. We also discuss the heat flux and its components as normalized by the saturation heat flux $Q_0 = \frac{3}{2} n_e k_B T_e \sqrt{2 k_B T_e/m_e}$, where we utilized the core density and temperature values to define this quantity. The saturation heat flux thus represents the heat flux corresponding to the full electron core thermal energy moving at the electron core thermal velocity with respect to the plasma frame. This normalization largely removes the radial variation of the heat flux. It also puts the heat flux in a convenient form in order to compare it with theoretical predictions of both collisionless and collisional heat flux regulation mechanisms, most of which utilize a similar normalization. For this case $Q_0 = 2.06 \times 10^{-2} \ \mbox{W/m}^2$, implying normalized parallel and antiparallel heat flux components of -0.0039 and -0.077, and a total normalized heat flux of -0.081.              

\section{Near-Sun heat flux}
\label{sec:hf}

\begin{figure*}
\sidecaption
    \includegraphics[width=12cm]{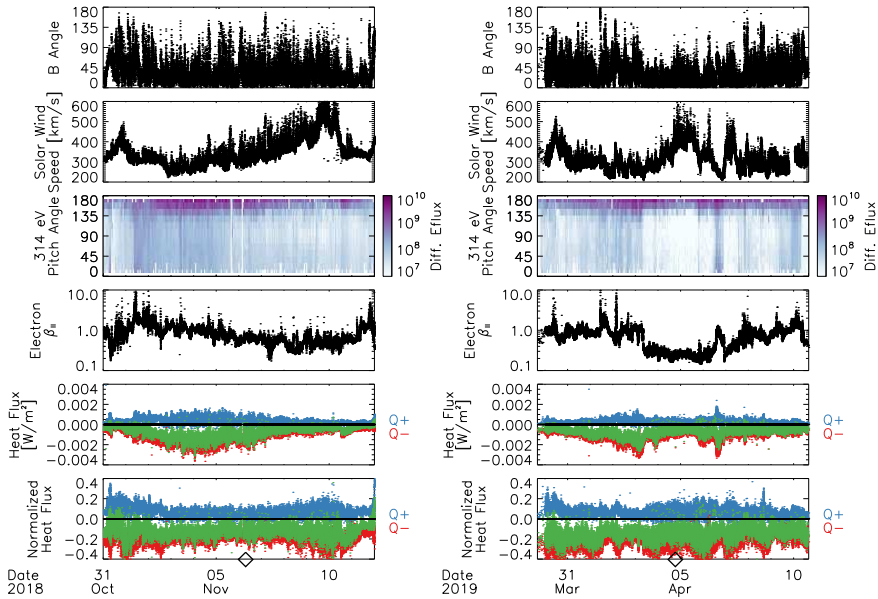}
    \caption{Solar wind parameters measured on the first (left column) and second (right column) PSP orbits. The panels show the angle between the magnetic field and the Sun direction, the solar wind speed (from SPC), electron pitch angle distributions at 314 eV, the ratio between the parallel electron thermal pressure and the magnetic field pressure $\beta_{||}$, the electron heat flux, and the electron heat flux normalized by the saturation heat flux. The net heat flux is shown in green, together with components corresponding to the portions of the electron VDF parallel (Q+, blue) and antiparallel (Q-, red) to the magnetic field. Diamonds on the time axes indicate perihelion on each orbit.}
    \label{fig:e12}
\end{figure*}

The first two PSP orbits remained entirely in the inward magnetic sector and did not cross the HCS during the near-Sun encounter \citep{badman_magnetic_2020, szabo_heliospheric_2020}. Figure \ref{fig:e12} displays a selection of relevant solar wind parameters for the first and second PSP orbits, for time periods near the perihelia (03:27 UT on 06 Nov 2018 and 22:40 UT on 4 Apr 2019). During these orbits, PSP encountered relatively low-beta plasma with ubiquitous Alfv\'{e}nic switchbacks near the Sun \citep{bale_highly_2019, kasper_alfvenic_2019}, easily visible as brief rotations in the magnetic field angle and associated increases in solar wind speed. The electron beta values remained mostly between 0.2 and 1 during the encounter periods, with only a few brief excursions to higher beta. The solar wind electron VDFs on these orbits contained a well-formed strahl that carried a relatively steady electron heat flux, with a magnitude generally larger at smaller heliocentric distances as expected, but also anticorrelated with solar wind speed \citep{halekas_electrons_2020}. The electron heat flux on these orbits was dominantly carried by the strahl and was unipolar in nature and reliably antiparallel to the magnetic field, as expected in the inward magnetic field sector. When divided by the saturation heat flux as described above, the resulting normalized electron heat flux values displays an apparent anticorrelation with beta for the majority of the two encounters. This persistent anticorrelation suggests the operation of collisionless heat flux regulation mechanisms, nearly all of which are expected to provide heat flux regulation that depends on beta.  

\begin{figure*}
\sidecaption
    \includegraphics[width=12cm]{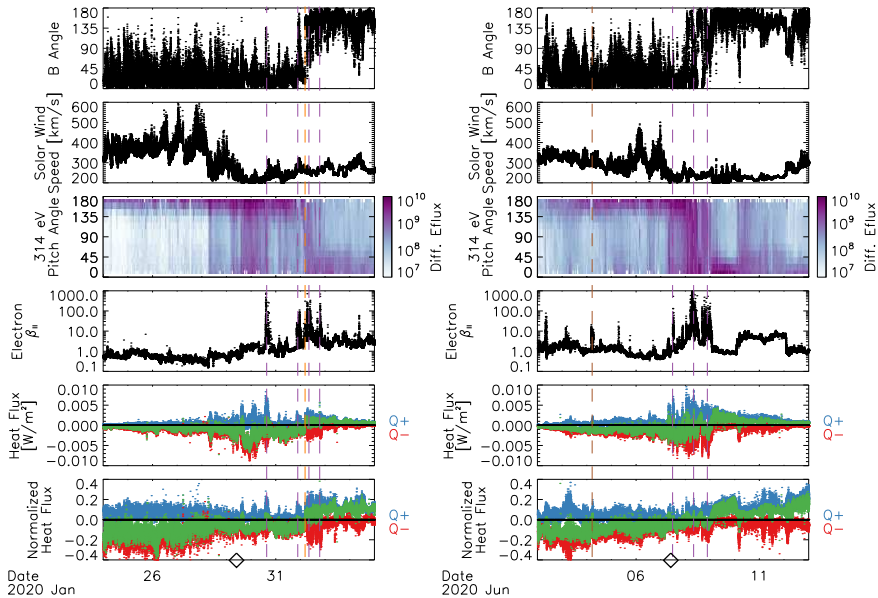}
    \caption{Solar wind parameters measured on the fourth (left column) and fifth (right column) PSP orbits, in the same format as Fig. \ref{fig:e12}, except that the solar wind speed is from SPAN-I. Dashed vertical lines indicate high-beta periods with bidirectional heat flux (purple), a true sector boundary (TSB, orange), and a potential disconnection event (brown).}
    \label{fig:e45}
\end{figure*}

The fourth and fifth PSP orbits took it closer to the Sun than its first three orbits, to a heliocentric distance of 0.125 AU. These orbits also differed from the earlier orbits in that they crossed the HCS near perihelion, rather than remaining in the inward polarity sector during the primary encounter as in earlier orbits \citep{badman_magnetic_2020, szabo_heliospheric_2020}. As a result, one cannot easily differentiate between the effects of heliocentric distance and those due to the very different plasma regimes encountered. Figure \ref{fig:e45} displays a selection of relevant solar wind parameters for the fourth and fifth PSP orbits, for time periods near the perihelia (09:37 UT on 29 Jan 2020 and 08:23 UT on 7 Jun 2020). The two encounters have many similar characteristics. Both inbound portions occur in the inward polarity sector, in regions with moderate solar wind speed and many switchbacks. Both inbound portions also contain intriguing quasi-periodic structures with time scales on the order of a day. Both orbits display a transition to lower speed and generally higher beta solar wind plasma. Both orbits cross from the inward to the outward polarity sector within a few days of perihelion. On the fourth orbit we observe a clear true sector boundary (TSB) consisting of a single sharp transition between antimagnetic field-aligned strahl and magnetic field-aligned strahl, colocated with a distinct HCS crossing. On the fifth orbit, the HCS and TSB are both apparently obscured by extended high-beta regions, potentially representing the HPS. Several similar but less extended regions occur before the HCS crossing on the fourth orbit. These high-beta regions primarily result from magnetic field decreases, but some also contain increases in density. Though they may represent transient structures, their occurrence on these orbits is surprisingly frequent.  

The net electron heat flux on the fourth and fifth PSP orbits generally increases in magnitude with decreasing heliocentric distance, as expected and as observed on the first two orbits. On the other hand, the normalized heat flux generally decreases in magnitude with decreasing heliocentric distance, and we observe an apparent anticorrelation with plasma beta similar to that observed in the first two orbits, as one might expect if beta-dependent collisionless mechanisms regulate the heat flux. 

The highest beta values on the fourth and fifth orbits appear in the regions tentatively identified as the HPS. In these regions, we also observe an anticorrelation between net electron heat flux and plasma beta. This could result from the same mechanism(s) that lead to the anticorrelation seen in other regions, or may instead result from the different physical origin and evolution of the plasma in these unique regions. Intriguingly, though the net heat flux consistently decreases in high-beta regions, the heat flux carried by the parallel and antiparallel portions of the VDF often has comparable magnitudes or in some cases increases in these regions as compared to the surrounding time periods. The electron pitch angle distributions also become much more isotropic at these times, implying that the omnidirectional suprathermal electron flux is also comparable to or higher than in the surrounding regions. We only interpret cases with a drop in both net electron heat flux and omnidirectional suprathermal flux as disconnections. This suggests that most observed heat flux dropouts near the Sun do not represent disconnections, but rather enhanced scattering, consistent with the observed bidirectional or isotropic suprathermal pitch angle distributions, and in accord with the conclusions of previous studies utilizing data from 1 AU \citep{crooker_suprathermal_2003, pagel_assessing_2005, pagel_understanding_2005}. We observe one clear exception on 4 Jun 2020, when both the net heat flux and its components drop (not readily apparent on the extended time scale of Fig. \ref{fig:e45}), suggesting a true disconnection event. 

\begin{figure*}
\sidecaption
    \includegraphics[width=12cm]{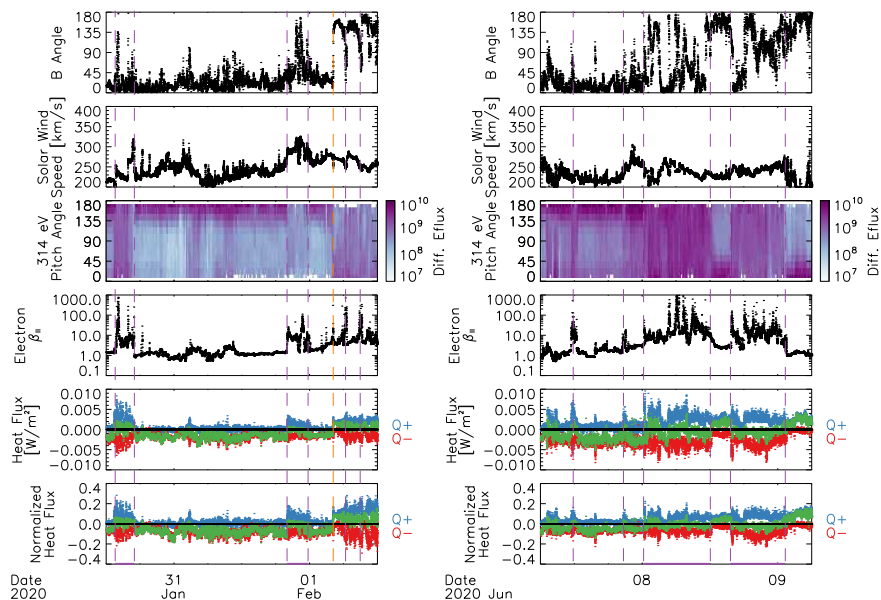}
    \caption{Expanded view of time periods near perihelion and the heliospheric current sheet. All panels are the same as in Fig. \ref{fig:e45}. Dashed vertical lines indicate high-beta periods with bidirectional heat flux or boundaries thereof (purple), and a TSB (orange).}
    \label{fig:e45_zoom}
\end{figure*}

Figure \ref{fig:e45_zoom} shows an expanded view of the time periods close to the HCS on both orbits, confirming the close correspondence between heat flux signatures and plasma beta, in agreement with observations from the earlier PSP orbits and also previous studies using data from greater heliocentric distances \citep{crooker_suprathermal_2003}. The very high beta regions may represent the HPS. However, several such periods occur well away from the main sector boundary crossing. Transient changes in the location of the HCS could explain these displaced encounters. However, an HPS clearly does not always surround the HCS near the Sun, given that the HCS crossing on the fourth orbit has no such surrounding high-beta region. An origin related to reconnection could explain the sporadic nature of the HPS \citep{sanchez-diaz_situ_2019, lavraud_heliospheric_2020}, and interchange reconnection could potentially explain the observations of HPS signatures far from the main sector boundary \citep{crooker_heliospheric_2004}. However, the remarkably extended duration (on the order of a day) of the two longest encounters on the fifth orbit appears difficult to explain through such transient mechanisms. Furthermore, most of the very high beta regions near the Sun, though they have low net heat flux, do not appear obviously consistent with disconnection, given the lack of a decrease (and often an increase) in the omnidirectional suprathermal electron flux.   

\begin{figure*}
\sidecaption
    \includegraphics[width=12cm]{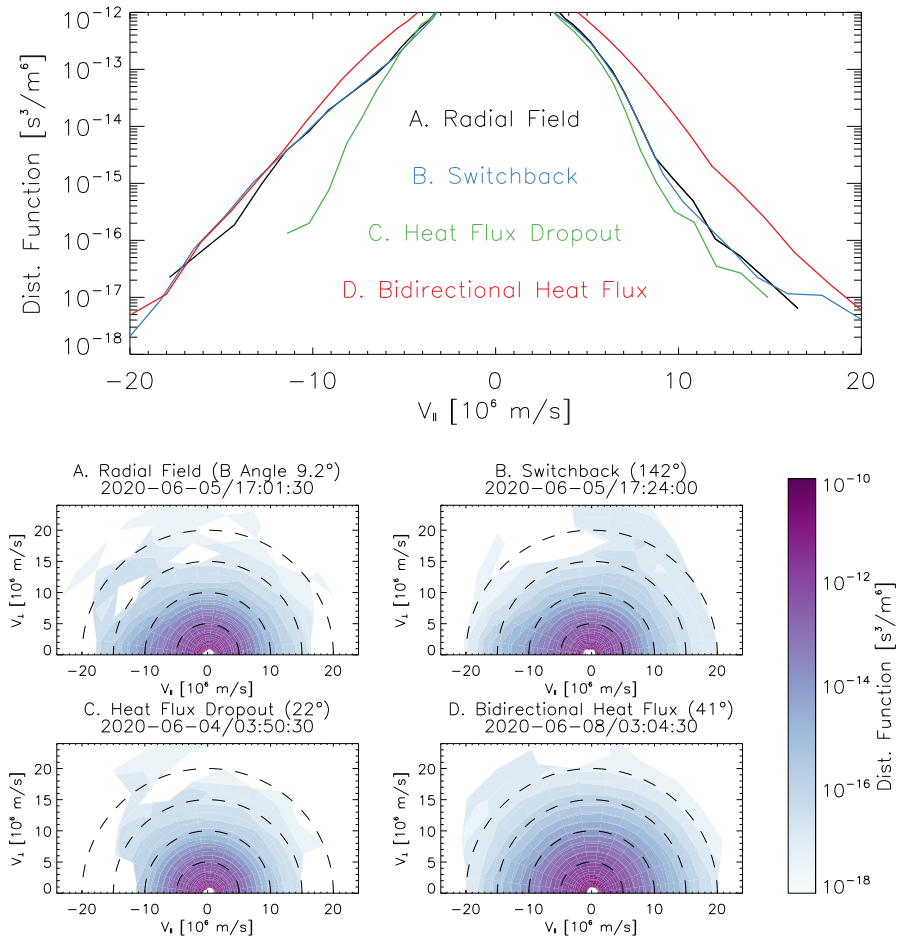}
    \caption{Electron VDFs in four different near-Sun regimes. The top panel shows cuts through the four VDFs along the magnetic field axis, and the bottom four panels each show a 2-d gyrophase-averaged VDF, in magnetic-field aligned coordinates.}
    \label{fig:vdfex}
\end{figure*}

We show four selected VDFs representative of plasma regimes encountered by PSP in Fig. \ref{fig:vdfex}. All four distributions have nearly isotropic core distributions, with at most slight parallel anisotropies. VDFs A and B (middle panels of Fig. \ref{fig:vdfex}) represent cases with a moderate net electron heat flux antiparallel to the magnetic field. These two observations occurred in quick succession, during a time period with moderate flow speed and plasma beta, on the inbound portion of the orbit. The two VDFs have very similar properties, despite almost inverted magnetic field directions during the observations. Both VDFs contain a clear antimagnetic field-aligned strahl, with an angular extent of $\sim 30-45 \degr$ (narrower at higher energies), and a tenuous and nearly isotropic halo outside of these angles. During switchbacks, as in VDF B, the electron heat flux reliably follows the magnetic field, allowing us to distinguish these magnetic field rotations from sector boundary crossings. 

VDFs C and D (bottom panels of Fig. \ref{fig:vdfex}) represent cases with very low net heat flux. VDF C comes from the potential disconnection event identified in Fig. \ref{fig:e45}. This VDF has low net heat flux, and also low heat flux components in both the parallel and antiparallel directions. The strong strahl in the antimagnetic field direction present in VDFs A and B is absent in VDF C, supporting the tentative identification of complete magnetic disconnection from the Sun. A weak halo contains suprathermal electrons with velocities primarily perpendicular to the magnetic field direction, with very little suprathermal flux in either the parallel or antiparallel direction. The observed perpendicular halo anisotropy could indicate a "magnetic bottle" geometry with higher magnetic field strengths in both directions along the field line.  

Finally, VDF D, from the high-beta region closest to perihelion (see Figs. \ref{fig:e45} and \ref{fig:e45_zoom}), also has very low net heat flux, but has large heat flux components in both the parallel and antiparallel direction. This VDF has a more significant halo with nearly isotropic suprathermal electron flux, larger in all directions than for the other three VDFs, with some indications of weak residual strahl-like populations parallel and antiparallel to the magnetic field. The former suggests that significant scattering has taken place, and the latter suggests the possibility of magnetic connection to the Sun at both ends of the magnetic field line. It is unclear how PSP could remain in a closed magnetic field geometry for an extended period of time. However, the high-beta regions might represent a mix of interspersed open and closed (or recently open and closed) magnetic field lines, as suggested by \citet{sanchez-diaz_situ_2019} and \citet{lavraud_heliospheric_2020}.    

\section{Heat flux organization}
\label{sec:org}

\begin{figure}
    \resizebox{\hsize}{!}{\includegraphics{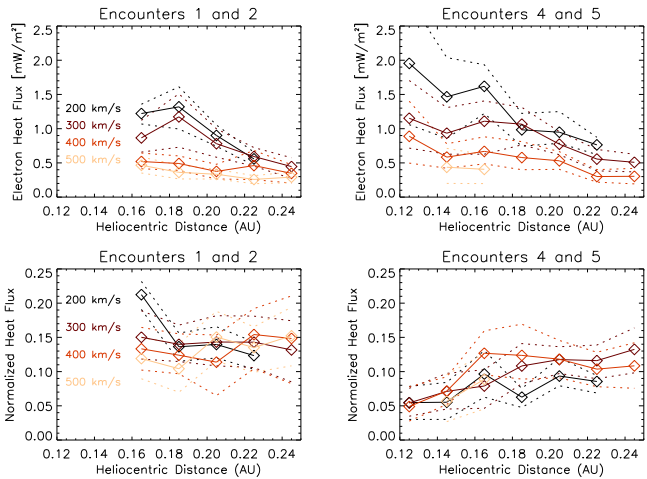}}
    \caption{Electron heat flux as a function of heliocentric distance and solar wind speed. The top panels show net heat flux magnitudes, and the bottom panels show normalized net heat flux magnitudes, for the first and second  (left panels) and the fourth and fifth (right panels) PSP orbits. The solid lines and diamonds show median values in five solar wind speed bins (each 100 km/s in width), and the corresponding dashed lines show the upper and lower quartiles.}
    \label{fig:hfvsw}
\end{figure}

While time series data suggest a relationship between plasma beta and electron heat flux, the heat flux (and its components) also has distinctive correlations with other solar wind parameters. We first investigate radial trends in the electron heat flux and its dependence on solar wind speed, as shown in Fig. \ref{fig:hfvsw}. We compare near-Sun observations from the first two PSP orbits (no HCS crossings, relatively low plasma beta) with those from the fourth and fifth orbits (HCS crossings, higher plasma beta). We do not consider the third orbit, which lacked solar wind velocity observations for a large portion of the orbit. For both data sets, we find that the electron heat flux generally decreases with increasing heliocentric distance in all speed bins. Furthermore, the net heat flux has a clear anticorrelation with solar wind speed, as reported by \citet{halekas_electrons_2020}. This trend, not clearly apparent at 1 AU \citep{salem_electron_2003}, is consistent with  expectations, given the lower heat flux carrying capacity of high-speed flows. The higher-speed wind generally has lower density everywhere, and also has a lower electron temperature near the Sun \citep{halekas_electrons_2020, maksimovic_anticorrelation_2020}, both of which lead to a lower value of the saturation heat flux. When we normalize the heat flux values by the saturation heat flux, the anticorrelation with solar wind speed largely disappears. The normalized heat flux also does not generally decrease with increasing heliocentric distance, and in fact increases with distance close to the Sun on the fourth and fifth orbits. This apparent radial trend on these orbits likely results from the presence of high-beta regions near perihelion. Indeed, all of the observed trends appear consistent with the operation of a beta-dependent collisionless heat flux regulation mechanism (or mechanisms) that limits the value of the normalized electron heat flux.  

\begin{figure*}
\sidecaption
    \includegraphics[width = 12cm]{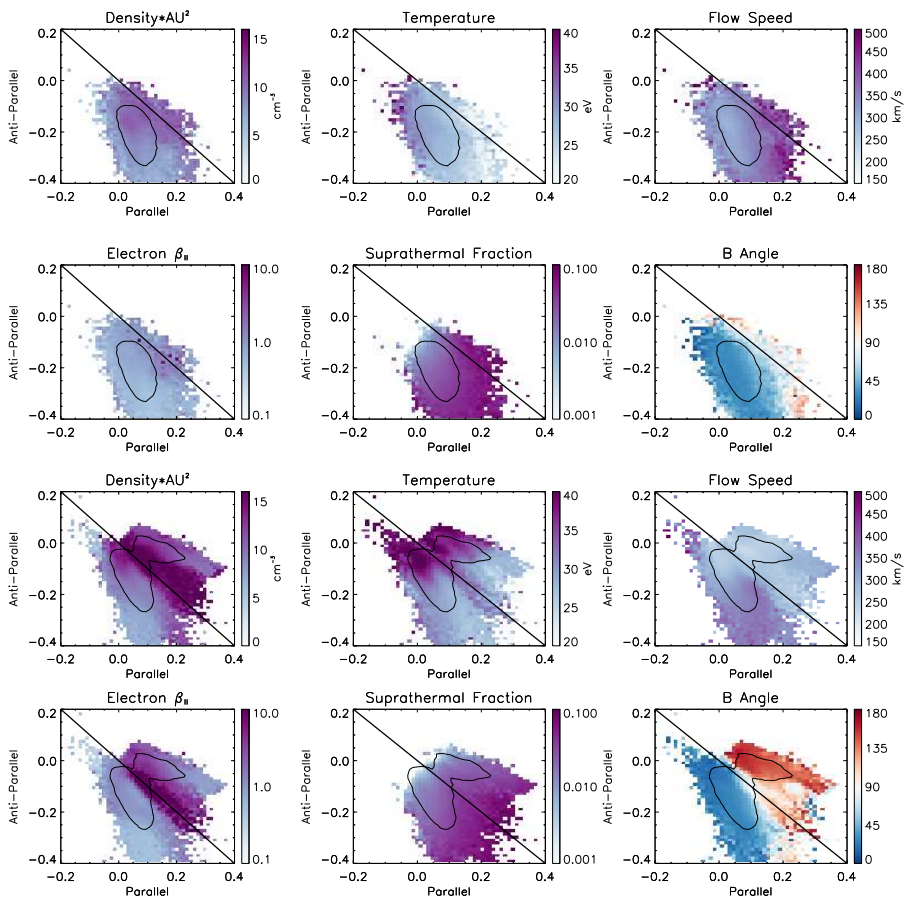}
    \caption{Solar wind parameters organized by electron heat flux. The top six panels show averages of solar wind density (normalized by $r^{-2}$ to remove radial trends), electron core temperature, flow speed, $\beta_{||}$, suprathermal fraction (suprathermal electron density divided by core electron density), and the angle between the local magnetic field and the Sun direction for the first two PSP orbits (heliocentric distances $<0.25$ AU), as a function of the normalized heat flux components corresponding to the portions of the electron VDF parallel and antiparallel to the magnetic field. The bottom six panels show the same quantities for the fourth and fifth PSP orbits. Diagonal lines mark equal and opposite parallel and antiparallel electron heat flux components. Contours enclose the region of parameter space with an occurrence frequency of more than 10\% of the maximum.}
    \label{fig:butterfly}
\end{figure*}

While the net electron heat flux has some trends with solar wind parameters, more trends appear when we separately consider the components of the heat flux corresponding to the parallel and antiparallel portions of the electron VDF. Figure \ref{fig:butterfly} shows selected solar wind parameters in a two-dimensional space formed by the parallel and antiparallel normalized electron heat flux components, for the same two data sets as in Fig. \ref{fig:hfvsw}. Electron VDFs with a more significant suprathermal electron population (higher suprathermal fraction) lie closer to the bottom right of the diagram, while those with nearly Maxwellian distributions lie closer to the origin. In this space, net heat flux dropouts show up along the diagonal line, with disconnections closer to the origin, and bidirectional or isotropic cases closer to the bottom right. In contrast, distributions with a dominant unidirectional strahl lie farther from the diagonal.  

Consistent with PSP's location in the inward polarity sector near the Sun during its first two orbits, the great majority of data points in the top six panels of Fig. \ref{fig:butterfly} lie below the diagonal line, with a larger antiparallel heat flux component, and thus a net antimagnetic field-aligned heat flux. Observations with a more unidirectional heat flux (strahl dominant over halo) lie farther to the left, and on average correspond to times with lower normalized density, higher electron temperature, lower flow speed, lower plasma beta, and more radial magnetic fields (the presence of switchbacks makes the average field less nearly radial then the expected Parker spiral direction). Observations with more equal heat flux components (halo dominant over strahl, heat flux dropouts, or bidirectional heat flux) correspond to times with higher normalized density, lower electron temperature, higher flow speed, higher plasma beta, and less radial magnetic fields. These time periods may represent regions closer to the HCS or HPS. 

Observations from the fourth and fifth PSP orbits follow most of the same trends. However, since these orbits crossed the HCS, points in the bottom six panels of Fig. \ref{fig:butterfly} lie on both sides of the diagonal line. The regions of parameter space close to the diagonal have much higher values of normalized density and plasma beta, corresponding to the high-beta regions near the HCS encountered on these orbits. The diagonal line lies close to the separatrix between observations with inward and outward average radial fields, as expected given the prevailing net electron heat flux outward from the Sun. As also apparent from Fig. \ref{fig:e45}, the outward polarity sector has higher normalized density, higher plasma beta, and lower average solar wind speeds than the inward polarity sector on these orbits.  

\section{Heat flux constraints}
\label{sec:const}

The observations discussed above demonstrate the clear anticorrelation between net electron heat flux and plasma beta, which appears consistent with the operation of beta-dependent collisionless heat flux regulation mechanisms. However, questions remain as to the relative importance of collisional and collisionless mechanisms, as well as the specific mechanism(s) at play.

\begin{figure}
    \resizebox{\hsize}{!}{\includegraphics{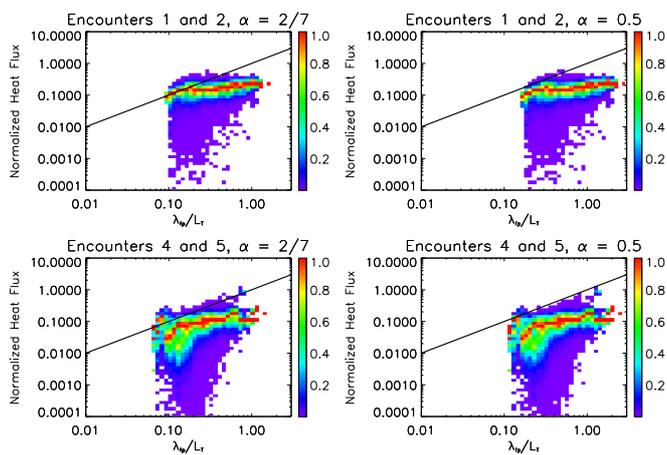}}
    \caption{Normalized electron heat flux as a function of a measure of collisionality. The top panels show 2-d frequency distributions of heat flux and the Knudsen number $K$ (the ratio between the mean free path $\lambda_{fp}$ and the temperature gradient scale $L_T$) for two values of the electron temperature exponent $\alpha$ (i.e. $T_e \sim T_0 r^{-\alpha}$), for the first two PSP orbits (heliocentric distances $<0.25$ AU). The bottom panels show the same quantities for the fourth and fifth orbits. Frequency values are column-normalized to a maximum of unity for each value of $K$, with the same color scale as \citet{bale_electron_2013} for ease of comparison. Diagonal lines mark a 1:1 relationship.}
    \label{fig:knudsen}
\end{figure}

Previous work \citep{salem_electron_2003, bale_electron_2013, horaites_self-similar_2015} has indicated that the electron heat flux observed at 1 AU follows the Spitzer-H{\"a}rm collisional limit or is bounded by it for a range of Knudsen numbers. In this limit, the normalized electron heat flux should be directly proportional to the Knudsen number. We repeated the analysis of \citet{bale_electron_2013}, who found that the electron heat flux at 1 AU matches the Spitzer-H{\"a}rm \citep{spitzer_transport_1953}  expectation for $K \la 0.3$ (more collisional plasma), but takes on a nearly constant normalized value of $\sim 0.3$ for $K \ga 0.3$ (more collisionless plasma). \citet{horaites_self-similar_2015} obtained similar results using Helios data from 0.3-1 AU. Figure \ref{fig:knudsen} shows a comparison for the same two data sets as in Figs. \ref{fig:hfvsw} and \ref{fig:butterfly}. Given the smaller temperature gradient scales close to the Sun, we have almost no data points for $K \la 0.1$. In the range of parameter space covered by the first two PSP orbits, we found heat flux values consistent with previous results \citep{bale_electron_2013}. The normalized heat flux has a nearly constant value of $\sim 0.1-0.3$ for $K \ga 0.2$, and decreases for the lowest sampled values of $K$, with values consistent with the Spitzer-H{\"a}rm limit for $0.1 \la K  \la 0.2$. However, this correspondence only holds for an electron temperature exponent $\alpha = \frac{2}{7}$, assumed by \citet{bale_electron_2013} because this value gives a constant conductive luminosity. For a larger exponent (which decreases $L_T$ and thus increases $K$) such as $\alpha = 0.5$, more consistent with observational results \citep{maksimovic_radial_2005, stverak_radial_2009, halekas_electrons_2020}, the normalized heat flux remains below the Spitzer-H{\"a}rm limit for all observed values of $K$. Meanwhile, though observations from the fourth and fifth orbits provide very little coverage of low Knudsen numbers, the great majority of the observed heat flux values lie below the Spitzer-H{\"a}rm limit for all parameter space sampled, regardless of the assumed temperature exponent. These results may prove consistent with simulations by \citet{landi_electron_2014}, who argued that, despite the apparent correspondence at 1 AU, the radial dependence of the electron heat flux does not follow the Spitzer-H{\"a}rm prediction for $K \ga 0.01$. To settle this question, one would have to consider the full radial variation of the heat flux. Regardless of the interpretation, the PSP observations appear to place the majority of the near-Sun environment firmly in the collisionless regime.  

\begin{figure}
    \resizebox{\hsize}{!}{\includegraphics{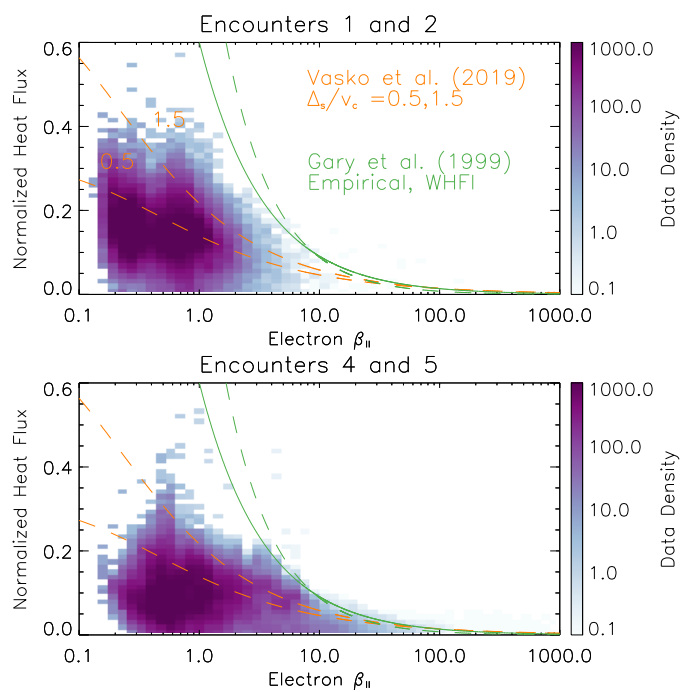}}
    \caption{Beta-dependent electron heat flux constraints. The two panels show 2-d frequency distributions of normalized electron heat flux and $\beta_{||}$ from heliocentric distances $<0.25$ AU from the first two PSP orbits (top panel) and from the fourth and fifth orbits (bottom panel). Orange lines indicate theoretical thresholds for the whistler fan instability \citep{vasko_whistler_2019} and green lines indicate an empirical threshold (solid) and a theoretical threshold (dashed) for the whistler heat flux instability \citep{gary_electron_1999}.} 
    \label{fig:vaskogary}
\end{figure}

The net electron heat flux in the near-Sun environment clearly depends on plasma beta, as discussed several times in the preceding sections. Figure \ref{fig:vaskogary} shows this result directly, for the same two data sets considered above. As described in Sect. \ref{sec:intro}, numerous collisionless heat flux mechanisms have been proposed. More than one of these could potentially provide the plasma beta-dependent heat flux regulation that our results suggest. Furthermore, multiple instabilities could contribute, either in concert, or in different plasma regimes \citep{shaaban_clarifying_2018, roberg-clark_suppression_2018}. In Fig. \ref{fig:vaskogary}, we compared the PSP observations to two theoretically predicted instability thresholds from the literature \citep{gary_electron_1999, vasko_whistler_2019}. Both predictions bound the observations well for both sets of orbits, particularly given the rather different plasma conditions encountered. Though many of the characteristics of the later orbits were very different from those of the earlier orbits, the overall organization of the net normalized electron heat flux by beta is very similar for the two data sets, notwithstanding the different ranges of beta sampled. The oblique whistler instability threshold \citep{vasko_whistler_2019} appears to better bound the majority of the observations for both data sets. Both of these predictions incorporate assumptions about the functional form of the electron VDFs, such as the thermal velocity of the strahl and the characteristics of the halo. The small fraction of the observations that lie at or above the predicted thresholds might therefore represent truly unstable plasma, or merely VDFs which do not satisfy the assumptions these calculations incorporated. Detailed calculations utilizing the actual measured VDFs could help determine which instability actually limits the electron heat flux in the near-Sun environment. 

\begin{figure}
    \resizebox{\hsize}{!}{\includegraphics{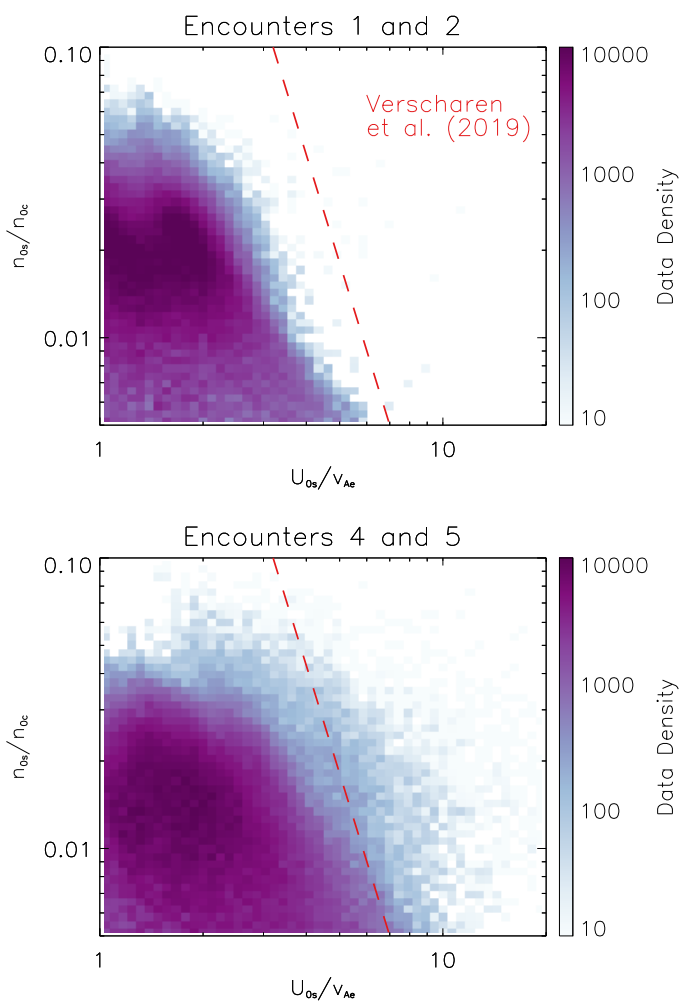}}
    \caption{Constraints on electron strahl parameters in the near-Sun environment. The two panels show 2-d frequency distributions of normalized strahl velocity and fractional strahl density from heliocentric distances $<0.25$ AU from the first two PSP orbits (top panel) and from the fourth and fifth orbits (bottom panel). The dashed red line indicates a theoretical threshold for an oblique fast magnetosonic instability \citep{verscharen_self-induced_2019}.} 
    \label{fig:verscharen}
\end{figure}

Another recent theoretical study \citep{verscharen_self-induced_2019} considered a similar oblique fast magnetosonic instability to that in \citet{vasko_whistler_2019} as a candidate to scatter the strahl. We compared the PSP observations to their calculations, by computing the fractional density and bulk velocity of the suprathermal electron population in the direction of the dominant heat flux component. Figure \ref{fig:verscharen} shows the results, together with the predicted instability threshold from \citet{verscharen_self-induced_2019}. We find that the predicted threshold completely bounds the observations from the first two PSP orbits (with room to spare), and bounds all but a tiny fraction of the observations from the fourth and fifth orbits. As discussed above, this could either indicate that a small fraction of the VDFs from the latter orbits represent unstable plasma, or that they do not satisfy the assumptions incorporated into the calculation. In either case, the PSP observations at or near the instability threshold present an interesting cadre for future study.  

\section{Conclusions}
\label{sec:conc}

The observations from PSP reveal a complex and structured near-Sun environment. The earlier PSP orbits, which remained in the inward sector and did not cross the HCS during the near-Sun encounter, encountered relatively low-beta plasma with ubiquitous Alfv\'{e}nic switchbacks near the Sun \citep{bale_highly_2019, kasper_alfvenic_2019}. The solar wind electron VDFs on these orbits contained a well-formed strahl that carried a relatively steady outward heat flux, with a magnitude anticorrelated with solar wind speed \citep{halekas_electrons_2020}. In contrast, the later PSP orbits, which crossed the HCS near the Sun, contain much more diverse electron heat flux signatures. On both the earlier and later PSP orbits, the normalized electron heat flux is anticorrelated with plasma beta, suggesting that this is a pervasive feature of the near-Sun solar wind. This global anticorrelation appears consistent with the operation of collisionless heat flux regulation mechanisms. The solar wind encountered on the later orbits spans a wider range of plasma beta. The very high beta regions tentatively identified as the HPS represent a different physical regime, and the anticorrelation there could result from different mechanisms. However, the relationship between heat flux and beta appears to follow the same trends, even in these very different regions. In any case, the apparent regulation of the heat flux explains the anticorrelation with solar wind speed on all orbits, since the higher-speed wind near the Sun has both lower density and lower electron temperature, leading to a lower saturation heat flux and thus lower heat flux carrying capacity.

The PSP observations appear inconsistent with the operation of purely collisional heat flux regulation mechanisms, since for a realistic temperature exponent the observations do not follow the Spitzer-H{\"a}rm limit for any range of Knudsen numbers observed, and therefore suggest a primary role for collisionless mechanisms. Many candidate mechanisms have been proposed, and multiple instabilities may play a role. However, theoretical predictions based on oblique whistler and magnetosonic instabilities provide thresholds that very closely bound the PSP observations of net electron heat flux and strahl parameters on all orbits to date. PSP measurements should provide the observations of whistler wave properties needed to confirm or refute the importance of such oblique instabilities.    

The very high beta regions encountered near the HCS, which may represent the HPS, contain electron VDFs that carry lower net heat flux. However, the parallel and antiparallel heat flux components, and the omnidirectional suprathermal electron flux, remain at a comparable level or even increase in most of these regions, seemingly inconsistent with disconnection from the Sun (with one exception). In a completely open diverging magnetic field geometry, it appears difficult for a single instability to force the electron VDF to evolve from one that contains a unidirectional strahl that carries a strong net electron heat flux to one that contains a bidirectional or isotropic suprathermal population that carries small or no net heat flux. The observed electron VDFs in these high-beta regions therefore may require the operation of multiple heat flux regulation mechanisms or a trapping mechanism such as a closed magnetic geometry or an electrostatic potential (i.e. a different or stronger potential than the normal interplanetary potential). In some high-beta regions, faint signatures of bidirectional electrons appear, possibly supporting the existence of such a trapping mechanism. However, the extended duration of some of the observed high-beta regions may argue against such a scenario. 

The PSP observations suggest that electron heat flux near the Sun is primarily regulated by collisionless plasma instabilities that depend on plasma beta. This has clear implications for how heat flows within and outward from the solar corona, and for the types of electron VDFs that can exist in different plasma regimes and at different heliocentric distances in our solar system. These results may also have broader implications for the nature of electron distributions and the associated electron heat conduction that occurs around other stars and in other astrophysical contexts, including accretion discs around objects such as black holes. 

\begin{acknowledgements}
We acknowledge the SWEAP contract NNN06AA01C for support. 
\end{acknowledgements}

\bibliographystyle{aa} 
\bibliography{references} 

\end{document}